# Periodic Structure with Electrostatic Forces: Interactions Beyond the Nearest Neighbor

**Sudesh Pathak**[*]
University of Mississippi
1764 University Circle
University, MS 38677
spathak@go.olemiss.edu

**Gagan Dangi**[*]
University of Mississippi
1764 University Circle
University, MS 38677
gdangi@go.olemiss.edu

**Farhad Farzbod**
Assistant Professor, University of Mississippi
203 Carrier Hall, University, MS 38677
farzbod@olemiss.edu

## ABSTRACT

*Periodic structures are a type of metamaterial in which the physical properties depend not only on the details of the unit cell but also on how unit cells are arranged and interact with each other. In conventional engineering structures, each unit cell interacts with adjacent cells. Methods developed for vibrational and wave propagation analysis in periodic engineering structures consider only nearest-neighbor interactions. The dispersion curves of such systems, in which only adjacent cells interact, have been extensively studied. Metamaterial properties depend on the interactions of a unit cell with other cells. Further interactions, and specifically, interactions beyond the closest neighbors, imply a more complex band structure and wave behavior. In this paper, an example class of such structures, in which electrostatic forces are the driving force, has been investigated. In this paper, properties affecting these periodic structures, such as elastic forces, have been investigated. An interesting property of such structures is that the band structures of such metamaterials can be tuned by changing electric voltages.*

## INTRODUCTION

Periodic structures have the interesting property that their vibrational characteristics depend not only on the materials and properties of their unit cell but also on how unit cells interact. In the vibrational analysis of these materials, Bloch's theorem [1] is commonly used. His work was followed by Mead et al. [2, 3], who developed a method for investigating harmonic wave propagation in periodic structures. Research has since taken place into a wide range of structures, such as mono-coupled periodic systems [4], beam grillage under harmonic point loading, [5] a periodic Bernoulli–Euler beam, [6], beam grids [7-9] and metamaterials with locally resonant components [10, 11]. There are many engineering applications of periodic structures, including wave beaming devices [12-14] and filters [1-3]. As well as linear periodic structures, nonlinear materials [15-19] and materials with dissipation [20-28] have been studied theoretically. In such work, the interaction between unit cells is limited to the nearest neighbor; each unit cell exerts forces on other unit cells with some boundary coordinates in common. This is because, in most engineering structures, the forces and interactions are limited to mechanical forces through contacts and joints. However, that is not the case for electromagnetic forces, such as the interatomic forces in a crystal. In crystals, for example, atoms interact with each other beyond the nearest neighbor. Magnetic and electrostatic forces have been used in periodic structures for modulation [29] and tuning [30], but they have not been used to design metamaterials with beyond-nearest-neighbor interaction. Chen et al. [31] investigated the winding number of the phase diagram in a 1-D chain when interactions extend to the second nearest neighbor. The present work is concerned with other aspects of electrostatic forces, the way dispersion curves are affected when the interactions reach beyond the nearest neighbor. Interactions beyond the nearest neighbor and their effects on the dispersion curves were first observed by Brillouin [1] in a 1-D chain and have been further studied by one of the authors [32, 33]. In a periodic structure, interactions running beyond the nearest neighbor have an effect on its dispersion curves. For example, the maximum number of wavevectors at each frequency in periodic materials depends on the various interacting layers [32]. It was also shown [32, 33] that dispersion curves are affected topologically when interactions are not limited to the nearest neighbor. In the following, we show that by using electrostatic forces, it is possible to fabricate such structures.

---

[*] These authors contributed equally

## STRUCTURES WITH ELECTROSTATIC FORCES

Consider a structure similar to the one depicted in Fig. 1. The slab and the columns are made out of a material that can be conductive such as copper, or non-conductive, such as glass or flexible Polyvinyl chloride (PVC) with a conductive coating. It can also be made by etching out a substrate to make rods and then coating them with conducting materials.

We can make all the rods connected to the same electrical voltage or columns with similar colors to the same electrical voltage, making the forces vary from repellant to the attraction with various amounts. Due to the nature of the electrostatic forces, the interactions in these structures run beyond the nearest neighbor. Also, the forces involved are nonlinear. However, suppose the structure is designed in such a way that the neutral position, is the equilibrium position. In that case, we can linearize the equations around the equilibrium position and use Bloch analysis. We first do force analysis and then investigate dispersion curves in the following subsections.

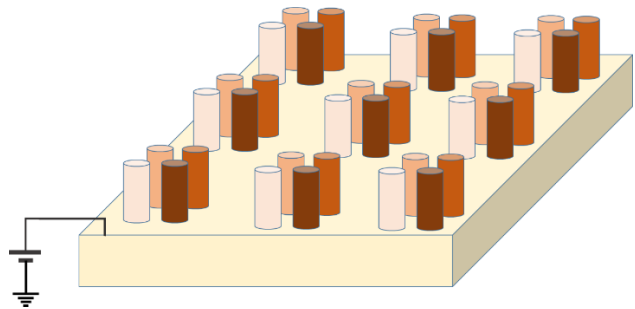

Figure 1: An example structure with electrostatic forces. Different colors represent possible different charges and/or materials.

**Force Analysis**

There are many possible configurations for a unit cell and how it is repeated to make the periodic structure. They can be arranged in a non-uniform shape or a circular one such as the ones in Fig. 2. However, we are considering a subset of these possibilities, in which columns are placed in a rectangular arrangement. To avoid confusion with matrix columns, we call the material columns masses from now on. It is also assumed that a unit cell has $m$ masses along the x-axis and $n$ masses along the y-axis. To simplify the formulation, we took that these masses are separated by distances $a$ and $b$ in the x and y direction (see Figure 3). The formulation is derived for the general case in which we have $m$ by $n$ number of columns in a unit cell. For the example structure of Fig. 3, $m=2$ and $n=3$. The reason for deriving equations for the general case is that we can then write a code for the general case and vary the design parameters to see their effect on the dispersion curves. This formulation is beneficial when considering hundreds of configurations and geometries for possible neural network training.

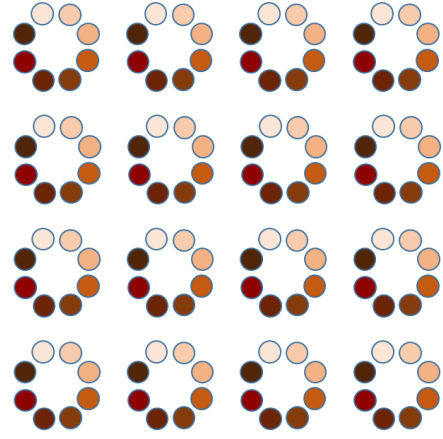

Figure 2: An example structure arranged in a non-rectangular geometry; similar to the elastic metamaterials, there are abundant possible configurations for a periodic structure with electrostatic forces.

We linearize the force equation for a generic unit cell located at $k$ and $l$, such as the one indicated by the double dashed line in Fig. 3. For the structure in Fig.3, the box with double dashed lines is at $(k,l)=(2,1)$, and for the reference unit cell with a solid dashed line $(k,l)=(0,0)$. Each mass in a unit cell is defined by two numbers such as $(r,s)$ where $r\leq m$ and $s\leq n$. Other geometrical parameters are depicted in Fig. 3. The distance between the masses at $(r,s)$ and $(u,w)$ in the x and y directions are:

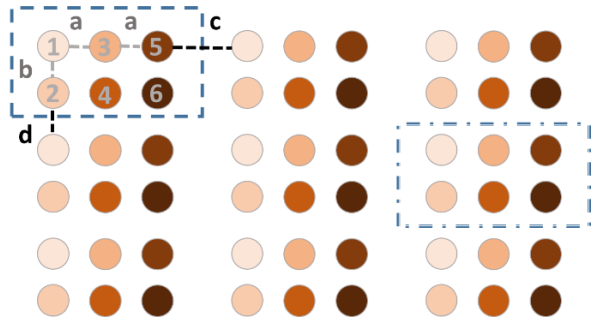

Figure 3: There are three columns and two rows in the depicted structure here. Distances between masses and unit cells in x and y directions are marked. Electrostatic forces between the reference unit cell and the one marked by double-dashed line is formulated.

$$d_x = k(m-1) \times a + (u-r) \times a + k \times c + x_{rs} - x_{uw} \quad (1)$$

$$d_y = l(n-1) \times b + (w-s) \times b + l \times d + y_{rs} - y_{uw} \quad (2)$$

In which $x_{rs}$ is the displacement of the mass $(r,s)$ of the reference unit cell from the equilibrium position in the x direction. Other displacements in the x and y directions are named similarly. Here, without loss of generality, we assume $x_{rs} > x_{uw}$, and $y_{rs} > y_{uw}$. We will take this assumption into account for the direction of the force when we write the equation of motion for each mass. This type of assumption is a common practice for simplifying the process of deriving equations of motion. The force between masses indexed by (r,s) and (u,w) in x and y directions are:

$$F_x(x,y) = \frac{K \times Q \times q}{d_x^2 + d_y^2} \cos\theta, \quad F_y(x,y) = \frac{K \times Q \times q}{d_x^2 + d_y^2} \sin\theta \tag{3}$$

in which Q and q are electric charges on the masses, K is the Coulomb's constant, and θ is the angle between the line connecting the centers of the columns and the x-axis. The vacuum's Coulomb constant equals $1/(4\pi\varepsilon_0)$, where $\varepsilon_0$ is the vacuum permittivity. For the air, it is very close to the one of vacuum. While we can use materials between masses with complex permittivity, we consider the air as the medium between masses. Using geometric values of sine and cosine, we can write:

$$F_x(x,y) = \frac{K \times Q \times q \times d_x}{(d_x^2 + d_y^2)^{\frac{3}{2}}}, \quad F_y(x,y) = \frac{K \times Q \times q \times d_y}{(d_x^2 + d_y^2)^{\frac{3}{2}}} \tag{4}$$

We want to linearize forces and stiffness relations such that:

$$F_x(x,y) = K_{xx} d_x + K_{xy} d_y, \quad F_y(x,y) = K_{yx} d_x + K_{yy} d_y \tag{5}$$

To simplify the process for linearization, we can denote $x_{rs} - x_{uw}$ and $y_{rs} - y_{uw}$ by x and y in both $d_x$ and $d_y$. Then find the partial derivative with respect to x and y. We evaluate the partial derivatives at equilibrium, in other words, when (x,y)=(0,0):

$$F_{x,x}(x,y) = F_{x,d_x}(x,y) \times d_{x,x} + F_{x,d_y}(x,y) \times d_{y,x} = F_{x,d_x}(x,y) \tag{6}$$

Using equations (1) and (2) to Evaluate $d_{x,x}$ and $d_{y,x}$ at (x,y)=(0,0), equation (6) takes the form of:

$$F_{x,x}(x,y) = KQq \left[ \frac{1}{(d_x^2 + d_y^2)^{\frac{3}{2}}} - \frac{3 \times (k(m-1) \times a + (u-r) \times a + k \times c)^2}{(d_x^2 + d_y^2)^{\frac{5}{2}}} \right] \tag{7}$$

$$F_{x,y}(x,y) = -KQq \left( \frac{3 \times k(m-1) \times a + (u-r) \times a + k \times c) \times l(n-1) \times b + (w-s) \times b + l \times d)}{(d_x^2 + d_y^2)^{\frac{5}{2}}} \right) \tag{8}$$

$$F_{y,x}(x,y) = F_{x,y}(x,y) \tag{9}$$

$$F_{y,y}(x,y) = KQq \left[ \frac{1}{(d_x^2 + d_y^2)^{\frac{3}{2}}} - \frac{3 \times ((l(n-1) \times b + (w-s) \times b + l \times d)^2}{(d_x^2 + d_y^2)^{\frac{5}{2}}} \right] \tag{10}$$

For an example, consider the structure in Fig. 3 in which *m*=2 and *n*=3. Forces along the x and y-direction between the masses located at (r,s)=(1,1), (u,w)=(2,2) and (k,l)=(2,1) can be found as:

$$F_x(x,y) = KQq \left[ \frac{1}{((6a + 2c)^2 + (2b + d)^2)^{\frac{3}{2}}} - \frac{3 \times (6a + 2c)^2}{((6a + 2c)^2 + (2b + d)^2)^{\frac{5}{2}}} \right] (x_{rs} - x_{uw}) -$$

$$KQq \left[ \frac{3 \times (6a + 2c) \times (2b+d)}{((6a + 2c)^2 + (2b + d)^2)^{\frac{5}{2}}} \right] (y_{rs} - y_{uw}) \tag{11}$$

$$F_y(x,y) = -KQq \frac{3\times(6a+2c)\times(2b+d)}{((6a+2c)^2+(2b+d)^2)^{\frac{5}{2}}} \times (x_{rs}-x_{uw}) + KQq\left[\frac{1}{((6a+2c)^2+(2b+d)^2)^{\frac{3}{2}}} - \frac{3\times(2b+d)^2}{((6a+2c)^2+(2b+d)^2)^{\frac{5}{2}}}\right](y_{rs}-y_{uw}) \quad (12)$$

As such, we get:

$$K_{xx} = KQq\left[\frac{1}{((6a+2c)^2+(2b+d)^2)^{\frac{3}{2}}} - \frac{3\times(6a+2c)^2}{((6a+2c)^2+(2b+d)^2)^{\frac{5}{2}}}\right] \quad (13)$$

$$K_{xy} = K_{yx} = KQq\left[\frac{3\times(6a+2c)\times(2b+d)}{((6a+2c)^2+(2b+d)^2)^{\frac{5}{2}}}\right] \quad (14)$$

$$K_{yy} = KQq\left[\frac{1}{((6a+2c)^2+(2b+d)^2)^{\frac{3}{2}}} - \frac{3\times(2b+d)^2}{((6a+2c)^2+(2b+d)^2)^{\frac{5}{2}}}\right] \quad (15)$$

Coefficient values obtained for the interaction of each mass of the reference unit cell with each mass of the interacting unit cell yield the stiffness matrix *K*. To find the eigenvalue problem, we follow the procedure explained in [33]. Each stiffness matrix is multiplied with the respective **T** and $\overline{\mathbf{T}}^T$ as linear push forward and backward operator and then the summation of these products along with the addition of $-\omega^2 M$ will yield an eigenvalue problem:

$$\left(-\omega^2\widehat{\mathbf{M}} + \sum_{\substack{k=0,\ldots p \\ l=0,\ldots q}} \overline{\mathbf{T}}_{(k,l)}^T \mathbf{K}_{(k,l)} \mathbf{T}_{(k,l)}\right)\widehat{\mathbf{q}} = \mathbf{0} \quad (16)$$

In this eigenvalue equation, (*k,l*) represents the position expressed as coordinates of the interacting unit cell with the reference unit cell, and $\widehat{\mathbf{q}}$ is the minimum set of displacements for the reference unit cell. We separate the x and y coordinate of the masses and stack them up on each other. Following this practice, for the example structure of Fig. 3 we get:

$$\widehat{\mathbf{q}} = \begin{bmatrix} x_{(0,0)_1} \\ x_{(0,0)_2} \\ x_{(0,0)_3} \\ x_{(0,0)_4} \\ x_{(0,0)_5} \\ x_{(0,0)_6} \\ y_{(0,0)_1} \\ y_{(0,0)_2} \\ y_{(0,0)_3} \\ y_{(0,0)_4} \\ y_{(0,0)_5} \\ y_{(0,0)_6} \end{bmatrix}, \mathbf{T}_{(2,1)} = \begin{bmatrix} \mathbf{I}_{12} \\ e^{i(2\mu_x+\mu_y)}\mathbf{I}_{12} \end{bmatrix}, \overline{\mathbf{T}}_{(2,1)}^T = [\mathbf{I}_{12} \quad e^{-i(2\mu_x+\mu_y)}\mathbf{I}_{12}] \quad (17)$$

In which $\mathbf{I}_{12}$ is the identity matrix with the Dimension of 12. Since all the masses interact with each other in both the x and y direction, $K_{(2,1)}$, the stiffness matrix for $(k,l)=(2,1)$ is a 24x24 matrix with no element being zero. Four of those elements are the ones in eq.'s 13-17.

**Dispersion Curves**

A code was written to vary various physical properties and design elements and plot the dispersion curves. A rectangular arrangement of masses similar to the one depicted in Figure 3 was assumed. Among possible configurations, for the ensuing investigation, we envisioned two-by-two mass columns in each unit cell. The columns are with radius and a height of 0.35 and 2 mm. We envisioned these columns to be made out of silicon, so the mass of each column is 1.793 milligram. The electric charge on each column is assumed to be 500 picocoulomb. Using notation similar to the ones in Fig. 3, for the current structure, we have distances a, b, c, and d equal to 1.5, 2.5, 2, and 3 mm, respectively. In this example, we considered interactions up to the eighth nearest neighbor. The following subsections further discuss the logic for a cut-off number of eight. We also investigate the effect of other parameters on the dispersion curves in these electrostatic structures. For the explained structure, dispersion curves are depicted in Fig. 4. There are eight branches representing two degrees of freedom for each of the four masses. There is a bandgap in the upper range of the band structure.

*Electric Charges*

The spring constants in eq. (13)-(15) are linearly proportional to the electric charges. Consequently, if the electric charges are multiplied by, for example, 10, then all the dispersion curves are scaled up by a factor of 10 in frequency. This property is, in fact, a significant advantage of such structures. Because electric charges can be changed by applying different voltages to the structure. As an example, the bandgap in Fig.4 can be tuned and changed easily. Also, the wave beaming can be adjusted quickly. This paper has discussed and analyzed only similar charges, positive or negative, on each mass. However, it is possible to design the structure with positive and negative charges. The arrangement of different charges should be so that we have a stable equilibrium position.

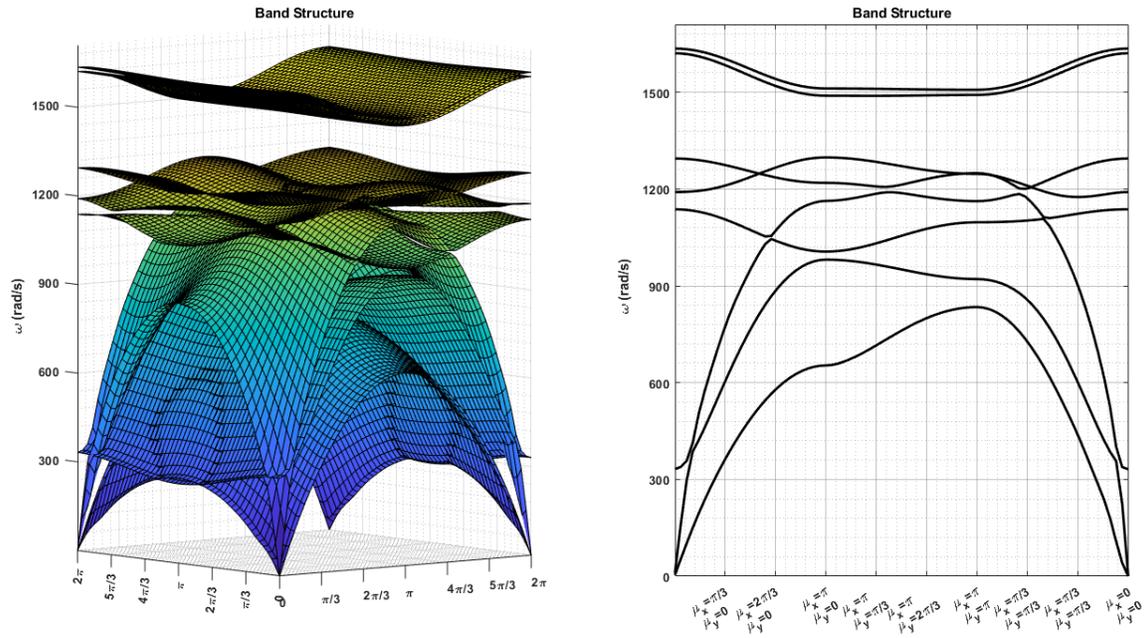

**Figure 4:** Dispersion curves of a structure with unit cell of four columns arranged in a two-by-two configuration. The columns are with radius and the height of 0.35 and 2 mm and the mass of 1.793 milligram. The electric charge on each column is 500 picocoulomb.

*Elastic Stiffness Effect*

In the stiffness calculations so far, only electrostatic forces have been considered. Inclusion of elastic forces due to the deflection of the beams is also required to model the stiffness completely. The inclusion of the elastic stiffness is not a difficult task, however. Beams of the proposed structures are not mechanically connected and behave similarly to a cantilever beam. As such, only the diagonal terms on the stiffness matrix have to be updated, namely the diagonal terms on the **K₀,₀** matrix. The stiffness term that should be added due to the distributed load can be stated as [34]; $8EI/L^3$ in which $E$ is the Young's modulus, $I$ area moment of inertia of cross section, and $L$ is the length of the beam. For a material such as copper with elastic modulus of 112 GPa, the mechanical stiffness relative to the electrical stiffness terms in **K_{k,l}** becomes a significantly bigger number. For the example structure that was used with columns of radius, height, and electric charge of 0.35mm, 2mm, and 500 picocoulomb, the elastic stiffness is about one hundred thousand times bigger than the terms due to electrostatic forces. On the other hand, for flexible PVC, the elastic modulus was measured to be 4.43 MPa [35]. In this case, the elastic stiffness terms are about a hundred times bigger. In both cases, the dispersion curves look similar if scaled accordingly (see Fig. 5). For the stiff material, dispersion curves are, in fact, very close to each other relative to their magnitudes. The bandgap is still there; however, Δω/ω ≈ 1.3 Hz /860 kHz. On the other hand, for the soft PVC, Δω/ω ≈ 55 Hz /7.2 kHz. In other words, using a soft material will provide a relative wider range for frequencies and band gaps. Granting that fabricating long rods with close proximity would be a challenge, It should be mentioned that increasing column height by a factor of ten will reduce the stiffness effect by one thousand.

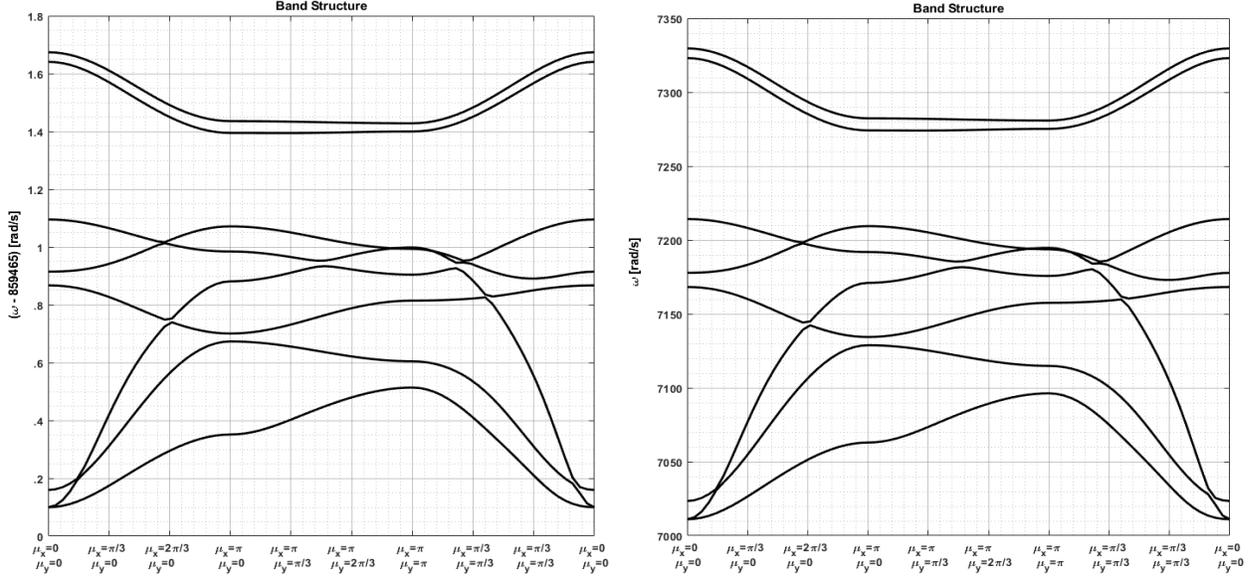

**Figure 5: Dispersion curves of two geometrically similar structure with one made with copper (depicted on the left) and the other with PVC. Frequencies on the left plot are subtracted by about 860 kHz.**

*Number of neighbors*

Electrostatic forces are inversely related to the square of charge distances. Nevertheless, in the proposed structure, the change in electrostatic forces around the equilibrium point is relevant, not the forces themselves. The deviation from the equilibrium position relative to the distances between masses is small such that we could linearize the forces and treat them like a linear spring. For the sake of argument, we consider forces along the vector connecting two masses separated by distance r in the x-direction. Linearizing this force means that the electrostatic interaction is replaced by a linear spring. This spring constant would be in the form of $\alpha \times 1/r^3$ for some constant $\alpha$. In other words, the stiffness goes down proportional to the reciprocals of the cubes of distances. Adding all spring constants along the x-direction, assuming masses are distanced equally, would amount to:

$$\beta \lim_{n \to \infty} \left( \frac{1}{1^3} + \frac{1}{2^3} + \frac{1}{3^3} + \cdots \frac{1}{n^3} \right), \tag{18}$$

in which β is a constant encompassing all coefficients such as electric charges, Coulomb constant, and distances between masses. This sum is in fact ζ(3) where ζ is the Riemann zeta function. The value of ζ(3) has been calculated up to more than a million digits, so it is simple to verify that the sum in (18) converges fast. For example, summing up to the n=8, the error goes down to about 0.5%. It should be mentioned that (18) only represents summation over one direction. The structure is considered to be stretched to infinity in two dimensions. As such, there will be infinite directions, and more summations similar to (18) have to be taken into account, so we have infinitely many summations. On the other hand, the sum starts from higher values of n and with increments of more than one for most directions. For this reason, even though not mathematically proven, we hypothesized that the convergence is still fast. We have investigated the convergence with a couple of examples. In Fig.4, dispersion curves are depicted for a sample structure up to the eighth nearest neighbor. For the same structure, interactions up to the sixth and tenth nearest neighbor are depicted in Fig. 6. As shown in Fig.6, dispersion curves overlap for low frequencies, and there are only some small differences for high frequencies.

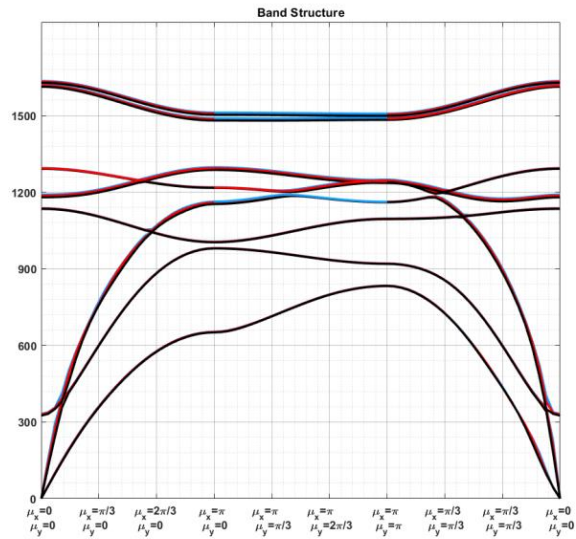

Figure 6: Dispersion curves for the structure of Fig. 4, when interactions up to the 6th (black lines), 8th (red lines) and 10th (blue lines) are considered. For low frequencies, they completely overlap, and for high frequencies, they are close.

**CONCLUDING REMARKS**

In this article, we investigated an example structure in which electrostatic forces are used as the means of energy transfer through mechanical vibrations. The difference between electrostatic forces and elastic forces is how they can reach beyond the nearest neighbor of each unit cell. Dispersion curves of such structures can be changed by varying electric voltage, making them a suitable candidate for tunable metamaterials. Dispersion curves of a couple of sample structures were investigated in this work.

# REFERENCES

[1] L. Brillouin, *Wave propagation in periodic structures: electric filters and crystal lattices*: Courier Corporation, 2003.

[2] D. J. Mead and S. Markus, "Forced vibration of a 3-layer, damped sandwich beam with arbitrary boundary conditions," *Journal of Sound and Vibration,* vol. 10, pp. 163-&, 1969.

[3] D. J. Mead, "Wave-propagation and natural modes in periodic systems .1. Mono-coupled systems," *Journal of Sound and Vibration,* vol. 40, pp. 1-18, 1975.

[4] M. G. Faulkner and D. P. Hong, "Free-vibrations of a mono-coupled periodic system," *Journal of Sound and Vibration,* vol. 99, pp. 29-42, 1985.

[5] R. S. Langley, N. S. Bardell, and H. M. Ruivo, "The response of two-dimensional periodic structures to harmonic point loading: A theoretical and experimental study of a beam grillage," *Journal of Sound and Vibration,* vol. 207, pp. 521-535, Nov 1997.

[6] D. L. Yu, J. Y. Fang, L. Cai, X. Y. Han, and J. H. Wen, "Triply coupled vibrational band gap in a periodic and nonsymmetrical axially loaded thin-walled Bernoulli-Euler beam including the warping effect," *Physics Letters A,* vol. 373, pp. 3464-3469, Sep 2009.

[7] G. Bordiga, L. Cabras, D. Bigoni, and A. Piccolroaz, "Free and forced wave propagation in a Rayleigh-beam grid: Flat bands, Dirac cones, and vibration localization vs isotropization," *International Journal of Solids and Structures,* vol. 161, pp. 64-81, 2019/04/01/ 2019.

[8] M. Botshekan, M. Tootkaboni, and A. Louhghalam, "On the dynamics of periodically restrained flexural structures under moving loads," *International Journal of Solids and Structures,* 2019/07/16/ 2019.

[9] A. Bacigalupo and M. Lepidi, "Acoustic wave polarization and energy flow in periodic beam lattice materials," *International Journal of Solids and Structures,* vol. 147, pp. 183-203, 2018/08/15/ 2018.

[10] M. Nouh, O. Aldraihem, and A. Baz, "Wave propagation in metamaterial plates with periodic local resonances," *Journal of Sound and Vibration,* vol. 341, pp. 53-73, 2015/04/14/ 2015.

[11] A. Ragonese and M. Nouh, "Prediction of local resonance band gaps in 2D elastic metamaterials via Bloch mode identification," *Wave Motion,* vol. 105, p. 102734, 2021/09/01/ 2021.

[12] A. H. Vonflotow, "Disturbance propagation in structural networks," *Journal of Sound and Vibration,* vol. 106, pp. 433-450, May 1986.

[13] Y. Yong and Y. K. Lin, "Propagation of decaying waves in periodic and piecewise periodic structures of finite length," *Journal of Sound and Vibration,* vol. 129, pp. 99-118, Feb 1989.

[14] C. Ma, R. G. Parker, and B. B. Yellen, "Optimization of an acoustic rectifier for uni-directional wave propagation in periodic mass–spring lattices," *Journal of Sound and Vibration,* vol. 332, pp. 4876-4894, 2013/09/30/ 2013.

[15] K. L. Manktelow, M. J. Leamy, and M. Ruzzene, "Weakly nonlinear wave interactions in multi-degree of freedom periodic structures," *Wave Motion,* vol. 51, pp. 886-904, 2014.

[16]   K. L. Manktelow, M. J. Leamy, and M. Ruzzene, "Analysis and Experimental Estimation of Nonlinear Dispersion in a Periodic String," *Journal of Vibration and Acoustics,* vol. 136, p. 031016, 2014.

[17]   K. Manktelow, R. K. Narisetti, M. J. Leamy, and M. Ruzzene, "Finite-element based perturbation analysis of wave propagation in nonlinear periodic structures," *Mechanical Systems and Signal Processing,* vol. 39, pp. 32-46, 2013.

[18]   B. P. Bernard, B. A. M. Owens, and B. P. Mann, "Uncertainty Propagation in the Band Gap Structure of a 1D Array of Magnetically Coupled Oscillators," *Journal of Vibration and Acoustics,* vol. 135, pp. 041005-041005-7, 2013.

[19]   B. P. Bernard, M. J. Mazzoleni, N. Garraud, D. P. Arnold, and B. P. Mann, "Experimental investigation of bifurcation induced bandgap reconfiguration," *Journal of Applied Physics,* vol. 116, p. 084904, 2014.

[20]   E. Andreassen and J. S. Jensen, "Analysis of Phononic Bandgap Structures With Dissipation," *Journal of Vibration and Acoustics,* vol. 135, pp. 041015-041015-8, 2013.

[21]   F. Farzbod and M. J. Leamy, "Analysis of Bloch's method in structures with energy dissipation," *Journal of Vibration and Acoustics,* vol. 133, p. 051010, 2011.

[22]   H. Al Ba'ba'a and M. Nouh, "An Investigation of Vibrational Power Flow in One-Dimensional Dissipative Phononic Structures," *Journal of Vibration and Acoustics,* vol. 139, pp. 021003-021003-10, 2017.

[23]   J. M. Manimala and C. T. Sun, "Microstructural design studies for locally dissipative acoustic metamaterials," *Journal of Applied Physics,* vol. 115, p. 023518, 2014.

[24]   A. O. Krushynska, V. G. Kouznetsova, and M. G. D. Geers, "Visco-elastic effects on wave dispersion in three-phase acoustic metamaterials," *Journal of the Mechanics and Physics of Solids,* vol. 96, pp. 29-47, 2016/11/01/ 2016.

[25]   F. Scarpa, M. Ouisse, M. Collet, and K. Saito, "Kirigami Auxetic Pyramidal Core: Mechanical Properties and Wave Propagation Analysis in Damped Lattice," *Journal of Vibration and Acoustics,* vol. 135, pp. 041001-041001-11, 2013.

[26]   A. Palermo and A. Marzani, "Limits of the Kelvin Voigt Model for the Analysis of Wave Propagation in Monoatomic Mass-Spring Chains," *Journal of Vibration and Acoustics,* vol. 138, pp. 011022-011022-9, 2015.

[27]   A. Aladwani and M. Nouh, "Strategic Damping Placement in Viscoelastic Bandgap Structures: Dissecting the Metadamping Phenomenon in Multiresonator Metamaterials," *Journal of Applied Mechanics,* vol. 88, 2020.

[28]   F. Farzbod, "Analysis of Bloch formalism in undamped and damped periodic structures," Ph.D. thesis, Georgia Institute of Technology, 2010.

[29]   M. Ansari, M. Attarzadeh, M. Nouh, and M. A. Karami, "Application of magnetoelastic materials in spatiotemporally modulated phononic crystals for nonreciprocal wave propagation," *Smart Materials and Structures,* vol. 27, p. 015030, 2017.

[30]   S. Chen, Y. Fan, Q. Fu, H. Wu, Y. Jin, J. Zheng*, et al.*, "A review of tunable acoustic metamaterials," *Applied Sciences,* vol. 8, p. 1480, 2018.

[31] H. Chen, H. Nassar, and G. Huang, "A study of topological effects in 1D and 2D mechanical lattices," *Journal of the Mechanics and Physics of Solids,* vol. 117, pp. 22-36, 2018.

[32] F. Farzbod, "Number of Wavevectors for Each Frequency in a Periodic Structure," *Journal of Vibration and Acoustics,* vol. 139, pp. 051006-051006-8, 2017.

[33] F. Farzbod and O. E. Scott-Emuakpor, "Interactions beyond nearest neighbors in a periodic structure: Force analysis," *International Journal of Solids and Structures,* vol. 199, pp. 203-211, 2020/08/15/ 2020.

[34] J. M. Gere and B. J. Goodno, *Mechanics of Materials* Cengage Learning; 8th edition 2012.

[35] F. Farzbod, M. Naghdi, and P. M. Goggans, "Using Liquid Metal in an Electromechanical Motor With Breathing Mode Motion," *Journal of Vibration and Acoustics,* vol. 141, pp. 014501-014501-4, 2018.